\documentclass[a4paper,11pt]{article}
\pdfoutput=1 % if your are submitting a pdflatex (i.e. if you have
\usepackage{jcappub} % for details on the use of the package, please
\usepackage[T1]{fontenc} % if needed

\emergencystretch=\maxdimen
\def\gw#1{gravitational-wave#1
  (GW#1)\gdef\gw{GW}}
\def\imbh#1{intermediate-mass black hole#1
  (IMBH#1)\gdef\imbh{IMBH}}
\def\gloc#1{Gravitational-wave Lunar Observatory for Cosmology#1
  (GLOC#1)\gdef\gloc{GLOC}}
\newcommand{\kj}[1]{{\textcolor{black}{{#1}}}}

\title{Gravitational-Wave Lunar Observatory for Cosmology }

\author[a]{K. Jani,$^{1,}$\note{Corresponding author.}}
\author[b]{A. Loeb}

\affiliation[a]{Department Physics \& Astronomy, Vanderbilt University, \\
2301 Vanderbilt Place, Nashville, TN, 37235, USA}
\affiliation[b]{Department of Astronomy, Harvard University, \\
60 Garden Street, Cambridge, MA 02138, USA}

\emailAdd{karan.jani@vanderbilt.edu}

\abstract{Several large-scale experimental facilities and space-missions are being suggested to probe the universe across the \gw{} spectrum. Here we propose \gloc{} - the first concept design in the NASA Artemis era for a GW observatory on the Moon. \kj{Using feasible interferometer technologies}, we find that a lunar-based observatory is ideal for probing \gw{} frequencies in the range between deci-Hz to 5 Hz, an astrophysically rich regime that is very challenging for both Earth- and space-based detectors. GLOC can survey binaries with neutron stars, stellar and intermediate-mass black holes to $\gtrsim 70\%$ of the observable volume of our universe without significant background contamination. The sensitivity at $\mathcal{O}(1~\mathrm{Hz})$ allows a unique window into calibrating Type Ia supernovae. \kj{At its ultimate sensitivity limits}, GLOC would trace the Hubble expansion rate up to redshift $z \sim 3$ and test General Relativity and $\Lambda$CDM cosmology up to $z\sim \kj{350}$.
}

\begin{document}
\maketitle
\flushbottom

\section{Introduction}
\label{sec:intro}

Observations from the first set of successful gravitational-wave (\gw{}) experiments - LIGO and Virgo - have shown the far reaching impact of the \gw{} spectrum from $10{\sim}1000$~Hz on fundamental physics, astronomy and cosmology \cite{Abbott_2019}. In the next two decades, \gw{} astronomy aim to probe the universe at lower frequencies and well beyond the sensitivity reach of the current detectors. The proposals for Earth-based next generation observatories include the Einstein Telescope \cite{Punturo:2010zz} and Cosmic Explorer \cite{CosmicExplorer}, which will constitute on the scale of tens of km intereferometers with enhanced sensitivity up to ${\sim} 5$~Hz. By early 2030s, space missions such as the Laser Interferometer Space Antenna (LISA) will open the \gw{} spectrum in the milli-Hz regime \cite{LISAmain}, while the global network of Pulsar Timing Arrays \cite{IPTA):2013lea} would be probing nano-Hz \gw{} frequencies. Other space-based concepts have also been proposed to deeper probe the milli- to micro-Hz frequencies \cite{Harry_2006, muARES, AMIGO}.

One of the most challenging spectral regimes to measure \gw{s} is from deci-Hz to 1~Hz. This frequency range tends to be too low for Earth-based detectors and too high for space missions. The universe offers a rich set of astrophysical sources in this regime \cite{Mandel_2018}, whose observations would lead to stringent tests of general relativity and physics beyond the Standard Model \cite{Sedda_2020}. A few proposals have been put forward for a space-based deci-Hz detector with geocentric (SAGE \cite{Lacour_2019}) and heliocentric (DECIGO \cite{2019IJMPD..2845001K}, ALIA \cite{Crowder_2005}, DeciHz Observatory \cite{Sedda_2020}) orbits, which rely on advanced technologies in the post-LISA era (2040s).   

In this study, we propose a \gw{} detector on the Moon whose primary goal is to access the deci-Hz range. With the advent of NASA's Artemis\footnote{\url{https://www.nasa.gov/specials/artemis/}} and Commercial Crew\footnote{\url{https://www.nasa.gov/exploration/commercial/crew/index.html}} programs, the time is ripe to consider fundamental physics experiments from the surface of the Moon. We find that the Moon offers an ideal environment for pursuing uninterrupted deci-Hz \gw{} astronomy for decades and will strongly complement with the Earth- and space-based network of telescopes. We suggest the acronym \textit{GLOC} - Gravitational-wave Lunar Observatory for Cosmology, for a detector that would survey $30-80\%$ of the observable volume of our universe for a wide-range of \gw{} sources. In particular, GLOC will be able to measure the evolution of the Hubble parameter at high-redshfits without multi-messenger followups.

\section{Gravitational-Wave Setup on the Moon}
The Moon offers a natural environment for constructing a large-scale interferometer as a \gw{} detector,  {and such a scenario have been mentioned earlier in the literature \cite{alma991023959399703276, 1993LPI....24..841L, potter_physics_1990}}. The atmospheric pressure on the surface of the Moon during sunrise is comparable to the currently implemented 8 km ultra high vacuum ($10^{-10}$ torr) at each of the LIGO facilities \cite{AdvLIGORef}. After sunset, the atmospheric pressure on Moon scales down to $10^{-12}$ torr \cite{Johnson1972}. The presence of vacuum just above Moon's solid terrain provides a great benefit in extending the interferometer length \kj{at minimal additional cost}. 

The seismometers left from the Apollo missions suggests that the Moon is much quieter than Earth (see \cite{Larose2005} and the references within). At low-frequencies \kj{($0.1 {\sim} 0,4$ Hz)}, the seismic noise on the Moon is \kj{almost three to four} orders of magnitude lower than on Earth \cite{Philippe_moon_seis, Hanada2005}. Seismic noise is a fundamental limitation for the low-frequency sensitivity of \gw{} detectors on Earth. In Advanced LIGO, the seismic noise dominates at frequencies $\lesssim 10$ Hz. The next generation Earth-based experiments are intended to push the limit to ${\sim} 3$ Hz through several ambitious improvements \cite{Evans:2016mbw} using quantum squeezers \cite{Goda2008}, cooling mirrors with a cryogenic system and building underground tunnels \cite{Hild_2009, Hild_2011}. By including some of these technological upgrades, GLOC can push the sensitivity to frequencies ${\sim} 0.1$ Hz (deci-Hz). These frequencies are too high to pursue with LISA-like space missions, as they are fundamentally limited by the quantum shot noise. 

\kj{Unlike a GW setup on earth, whose duty-cycle is routinely impacted by environmental factors (earth-quakes, lighting, winds), the lunar-based detector will have relatively fewer disturbances.}
The detector will be very mildly sensitive to the gravitational pull from the Earth's ocean waves. 
The Moon-quakes occur at much lower frequencies \cite{Hanada2005} and should not impact the \gw{} sensitivity in the GLOC spectrum, \kj{albeit thermal quakes and micrometeorite may impact observations}. Bombardment by cosmic rays and solar flares can be a source of non-Gaussian noise, which may require a certain magnetic shield around the end stations of the detector to keep the excess charge grounded. 
In addition, a cooling source can maintain a steady temperature and mitigate thermal expansion, as the temperature on the surface of the Moon changes from $-130^\circ$~C to $120^\circ$~C in a day \cite{Williams2017}. Alternatively, a few km long open lava tubes found on the Moon \cite{Chappaz2017} can provide a natural infrastructure for setting up the intereferometer, \kj{though will demand different geometry than equilateral triangle.} Such caves can also can protect from the lunar regolith that may cause light scattering.

An additional advantage (thus far) is that the Moon is not corrupted by any unpredictable noise from human activities. The site selection for the detector should avoid terrain favorable for potential launches.  In case of a lock-loss in the interferometer, a lunar-based detector can be brought back online from a control center on Earth. In the event of a serious hardware failure, parts of the detector can be replaced and repaired by astronauts. The benefit of performing \kj{any unforeseen} maintenance is not available for space-based \gw{} detectors, making the Moon a better long-term \kj{observatory}. In addition, future space-missions to access the deci-Hz range are limited in their lifetime (typically a few years), after which the gravitational perturbation from solar system objects will disrupt their geometry. In contrast, a lunar-based detector can operate and be steadily improved for decades. Because the \kj{key science} of the lunar-detector is at low frequencies, the data transmission rate is only a magnitude more than that expected for the space-mission like LISA.

\begin{figure*}
\centering
 \includegraphics[scale=0.22,trim = {0 20 0 0}]{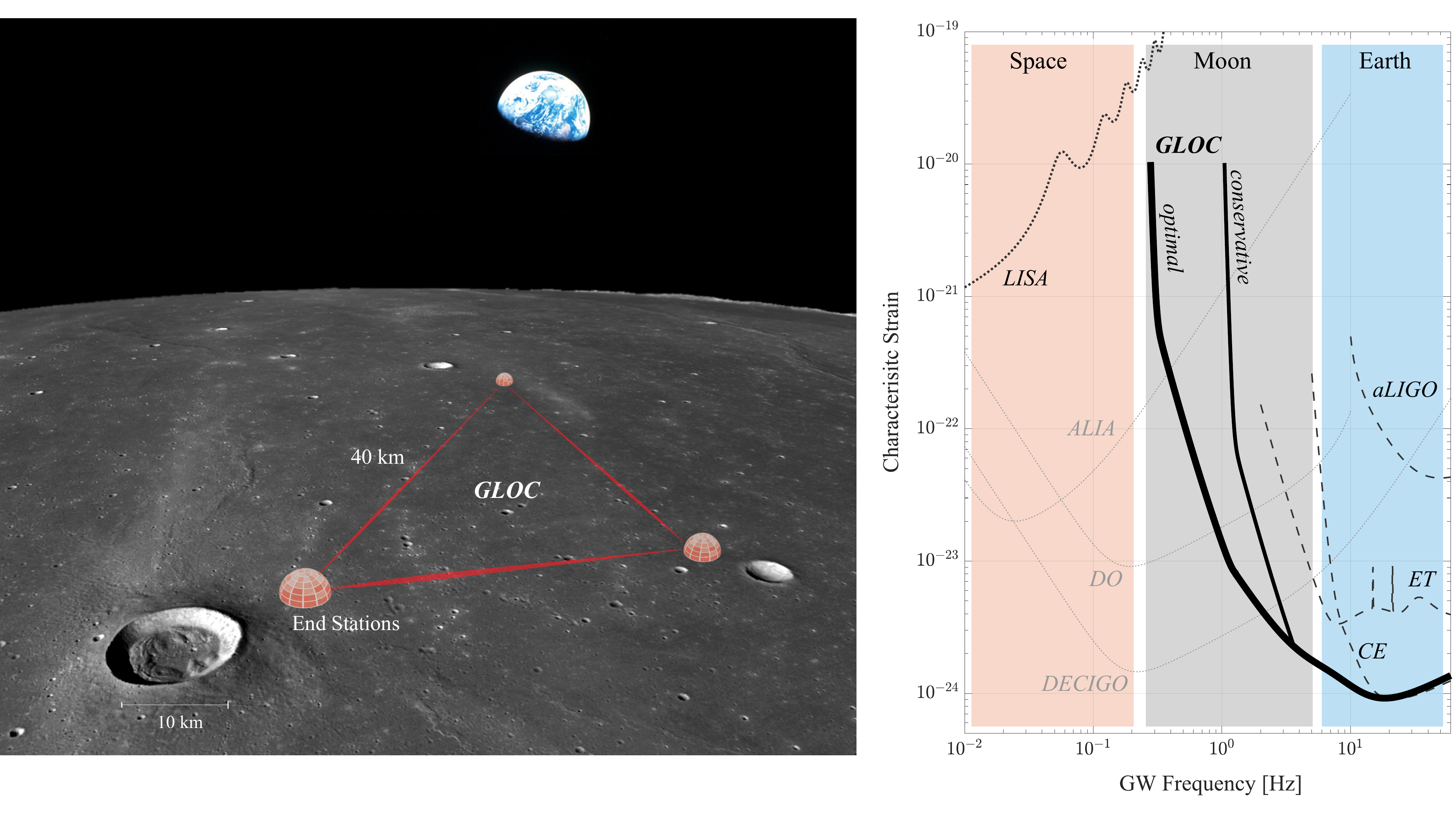}
\caption{{\bf Concept design for GLOC and predicted sensitivity.} Left: Three end stations on the surface of the Moon forming the full triangular-shape \gw{} detector. The end stations are separated by 40 km. Each end station will contain a test mirror and a laser, making GLOC equivalent of three interferoemters. Right: \gw{} sensitivity (as characteristic strain) of space-based (LISA), Earth-based (aLIGO, ET, CE) and lunar-based (GLOC-optimal and conservative) detectors.
Also plotted for comparison are sensitivity curves of space-based deci-Hz concepts DECIGO, ALIA, and DeciHz Observatory (DO). Image of Moon's surface was adapted from Lunar Reconnaissance Orbiter (NASA/GSFC/ASU) and Earthrise from the Apollo archives. }
\label{fig:concept} 
\end{figure*}

\section{GLOC Concept Design}  {To showcase the maximum potential for a lunar detector, we adopt a design schematic similar to that of next-generation Earth-based detectors Cosmic Explorer (CE) and Einstein Telescope (ET). As shown in the left panel of Fig. \ref{fig:concept} (left), the arm-length of a potential interferometer could be set to 40 km (or maximum attainable within the selected site), and the L-shaped LIGO type intereferoemter can be replaced by a triangular geometry.}  
Each end station forming the triangle can host the laser and two test masses. Therefore, the full GLOC could be equivalent of three independent detectors. These end stations can be designed as dorm shaped compartments that are temperature controlled and isolated from the rest of the detector. The curvature of the Moon leads to a ${\sim} 450$~m vertical offset for the light path between 40 km separation.  An ideal site for \gloc{} would be within a bigger craters ($\gtrsim20$~km), providing a flat land for at least two end stations and a higher elevation to place the third station. \kj{Another potential scenario is to put each station on the few km tall mountains.} Alternatively, a smaller optical setup can also be implemented within three landers separated by a few tens of km, though that would unlikely host the sophisticated technologies expected in ET and CE.

The right panel of Fig. \ref{fig:concept} shows the target sensitivity of GLOC versus \gw{} frequencies and compares it with other proposed detectors \cite{Robson_2019, Hild_2011}. We consider two projections for GLOC - \kj{the ultimate detector sensitivity} down to $f_\mathrm{low}=0.25$ Hz (GLOC-optimal), and a \kj{lesser sensitive version} that reaches $f_\mathrm{low}=1$ Hz (GLOC-conservative). The noise curves for GLOC can be downloaded from {\url{https://doi.org/10.5281/zenodo.3948466}}.

In both the stated cases of GLOC, we assume that the primary limiting noise in the mid to higher frequencies will be dictated by the quantum noise. 
%In our study, we compute this quantum noise using the CE2-ifo available on \texttt{pygwinc} \cite{pygwinc}.
Below $\mathcal{O}(1)$Hz, the sensitivity of GLOC is governed by the seismic noise and the suspension thermal noise. %To estimate the behavior of these noises in GLOC, we adapt their power-laws from the ET-D design. 
We expect the seismic noise in GLOC would be at least three orders of magnitude lower than an ET-like configuration on Earth. Due to the Moon's lower surface gravity, the noise from a suspension setup similar to that on earth would be reduced by a factor of ${\sim}3$. While the thermal noise will improve by implementing mirror coating \cite{PhysRevLett.122.231102}, \kj{future studies will provide a more accurate noise budget}.
%With these improvements, GLOC would achieve the projected conservative sensitivity. 
To reach the \kj{ultimate} sensitivity will require an unconventional suspension setup. A possible mechanism it to let the test masses be in a free fall with a so-called juggled interferometer \cite{2014CQGra..31x5006F}, \kj{though that may require diluting the high-frequency sensitivity for ET-like design}. The juggled interferometer setup is more favorable to implement on the Moon due to the freedom with the atmospheric vacuum.

\section{Methods} For constructing the sensitivity curve of GLOC, we apply the quantum noise in the range $2-5000$ Hz from the CE2-ifo available on \texttt{pygwinc} \cite{pygwinc}. To estimate the behavior of seismic and suspension noise below ${\sim2}$ Hz, we adapt their power-laws from the ET-D design and scale them to the improvement stated in the concept design. We compute the horizon distance of GLOC for various astrophysical binaries, characterized by their masses ($m_{1}, m_{2}$), using the methods described in Ref. \cite{2020NatAs...4..260J}. For the space-based detectors, we set a timeline of 4 years. For the \gw{} signal $h(f)$, we utilize the state of the art \gw{} signal model \texttt{IMRPhenomPHM} \cite{London_2018}. This model includes the radiated higher harmonics beyond the quadruple term that are crucial to probe intermediate-mass black hole binaries \cite{PhysRevD.97.024016}.

We set GLOC's detection threshold at signal to noise ratio (SNR) of 8. The lower limit of integration for SNR is started from $f_\mathrm{low}=0.25$ Hz and includes a factor to account for the three detector within GLOC's triangular geometry. We transform the luminosity distance into redshift through Planck 2018 Cosmology \cite{2018arXiv180706209P}.
To estimate the sky-localization error $\Delta{\Omega}$, we first compute the frequency bandwidth $\sigma_f$ and timing accuracy $\sigma_t$ for a given astrophysical source \cite{Fairhurst_2011}. The effective baseline $L$ is calculated through the orbital path of the Moon during the time spent by the \gw{} signal in GLOC's band. As the binaries spent most time at the low-frequency regime of GLOC, we approximate the angular uncertainty as $\sigma_\theta \sim \sigma_t (c /L)$ \cite{Lau_2020}.

\begin{figure}[t!]
\centering
 \includegraphics[scale=0.4,trim = {50 0 0 0}]{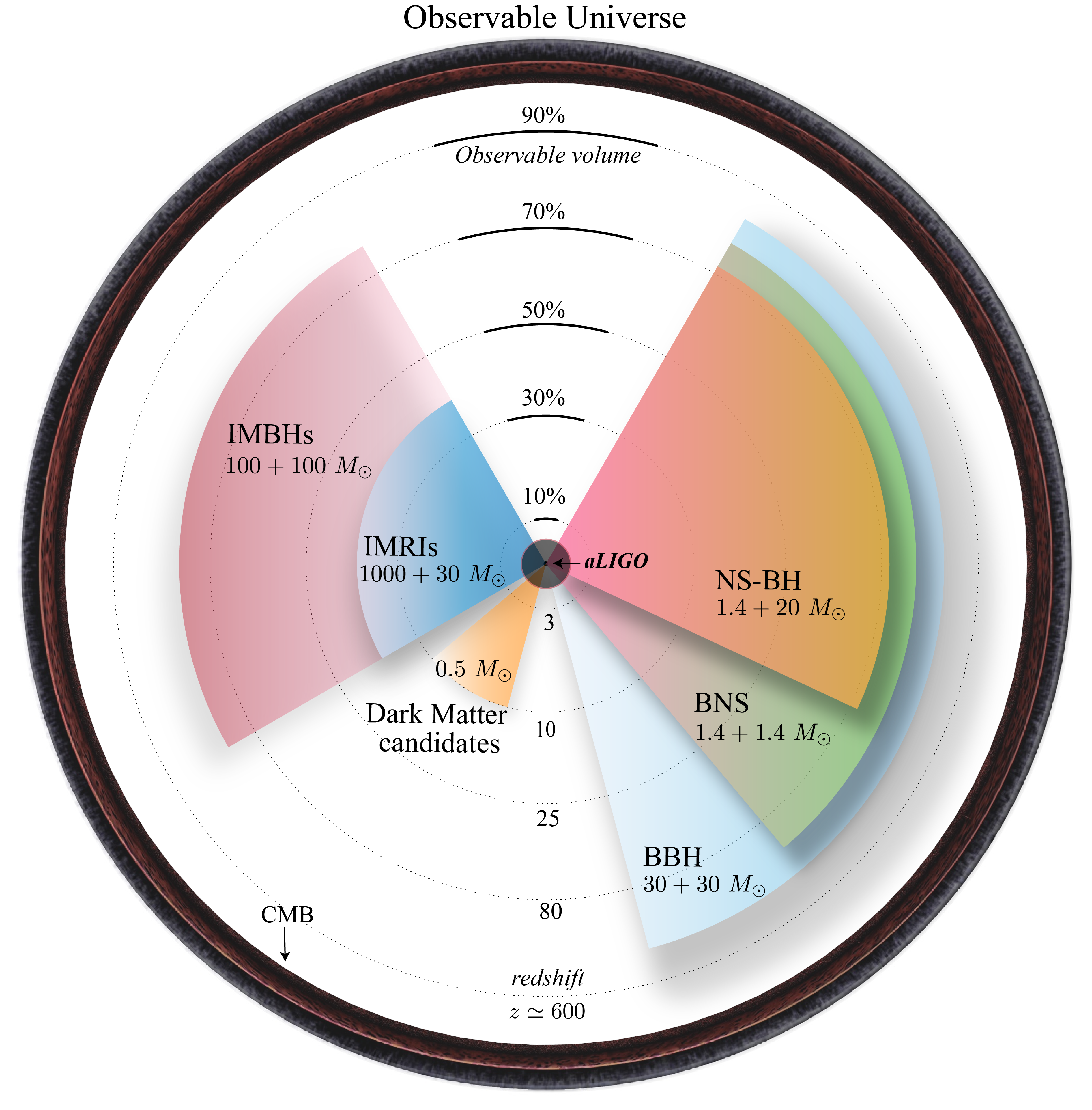}
\caption{{\bf Cosmological reach of GLOC in comoving coordinates.} The concentric circles represents the percentage fraction of the comoving volume of the observable universe $(V_\mathrm{obs} = 1.22 \times 10^4~\mathrm{Gpc}^3 $) out to a given cosmological redshift, with the outermost being the CMB \cite{2018arXiv180706209P}. The highlighted slices refer to the horizon redshifts in GLOC   {(optimal)} for the corresponding \gw{} sources at their detection threshold (SNR $\geq8$). The ones on the right are known examples of binary sources with black holes (BHs) and neutron stars (NSs), while the left exhibits potential discoveries of intermediate-mass black holes (IMBHs) and intermediate mass-ratio inspirals (IMRIs). For reference, the circle in the center represents the maximum reach of aLIGO at its design sensitivity \cite{ObservingScenarios}. }
\label{fig:reach} 
\end{figure}

\section{Science Case of GLOC} 
\begin{figure}
\centering
 \includegraphics[scale=0.22,trim = {50 150 0 220}]{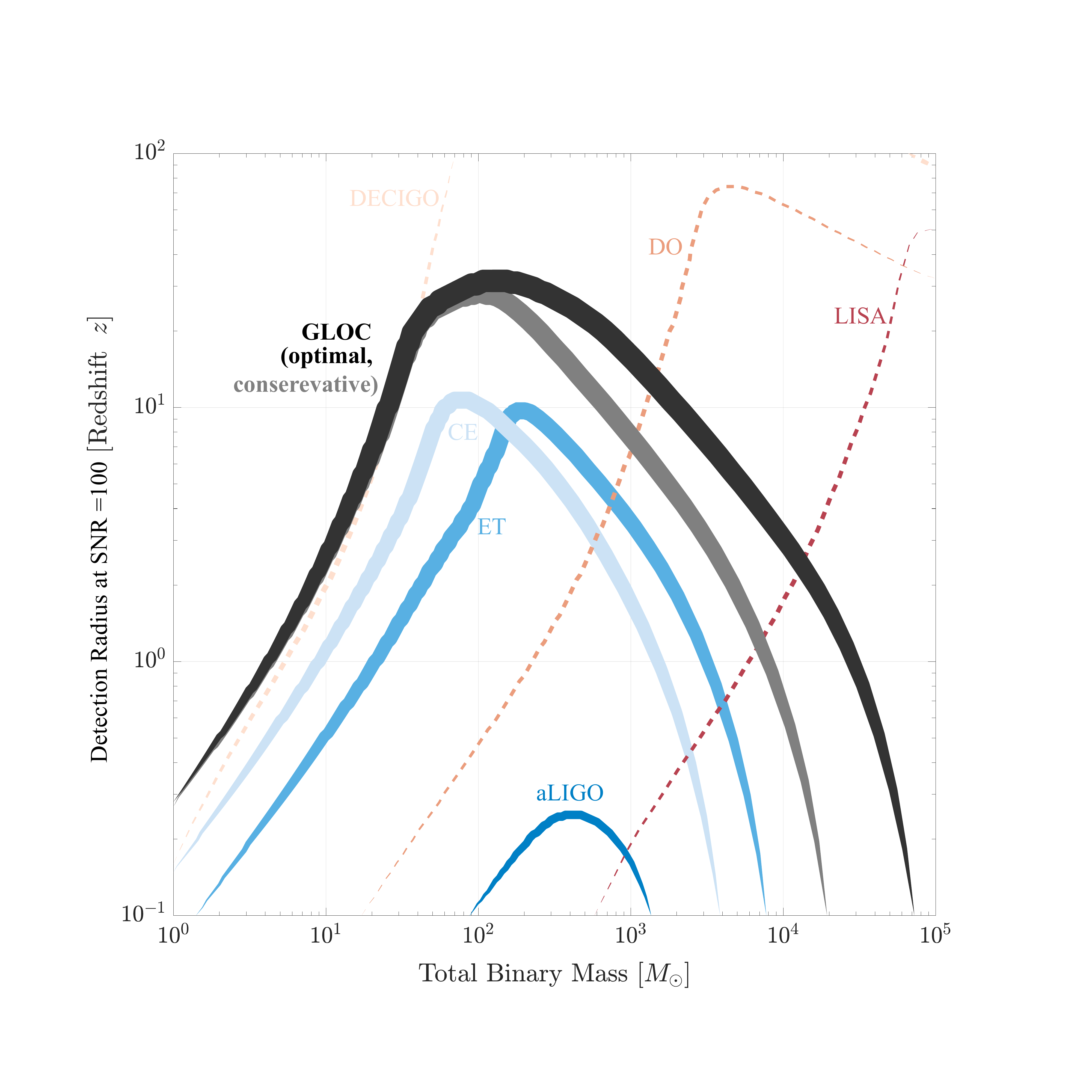}
\caption{{\bf High SNR detections.} The area under each line refers to the detection radius of equal-mass binaries at SNR $\geq100$ for different GW detectors. The thick black (optimal) and grey (conservative) lines refer to sensitivity of GLOC. 
The dashed lines refer to space-mission and colored lines indicate the ground-based detectors discussed in this text.  }
\label{fig:hor} 
\end{figure}

As showcased with fig. \ref{fig:reach}, the detector would have a rare advantage of accessing \gw{s} at cosmological distances across five orders of magnitude in mass - from sub-solar dark matter candidates $({\sim}10^{-1}\ M_\odot)$ \cite{Shandera_2018} to stellar mass binaries $({\sim}10^{1-2}\ M_\odot)$ to intermediate-mass black holes (IMBHs, ${\sim}10^{3-4}\ M_\odot)$ \cite{2006ApJ...653L..53A}. Across this entire mass-range, GLOC's sensitivity would outperform that of the upcoming \gw{} experiments on Earth (CE, ET) and space (LISA) (see fig. \ref{fig:hor}). Furthermore, the sensitivity band of GLOC is not expected to have any astrophysical foregrounds from the white dwarf binaries \cite{Robson_2019}. Thus, any \gw{s} with redshifted frequency $f_\mathrm{det} = (1+z)f_\mathrm{src} \gtrsim 0.2$~Hz could be identified in the data without contamination.

At the detection threshold (SNR$~ \geq 8$), we find that GLOC can detect the mergers of neutron stars (NSs) and stellar black holes (BHs) to over 70\% of the entire observable volume of our universe ($z {\lesssim} 100$). This provides an unprecedented cosmological probe of the early universe, comparable to the 21 cm cosmology proposed from the far-side of the Moon \cite{burns2012astrophysics} and the cosmic microwave background (CMB) experiments. While we do not expect stellar objects to exist beyond $z{\sim} 70 $ \cite{Avibook}, even one such detection will violate $\Lambda$CDM cosmology \cite{Koushiappas_2017}. The measurements of stellar binaries at such high redshifts would also constrain the earliest stellar population \cite{Jani_2020}.

With the low-frequency limit up to ${\sim} 0.2$ Hz, GLOC will start detecting the inspiral of stellar binaries a few months to days before their merger \cite{Peters1964}. This makes GLOC an early alert system for all coalescing \gw{} sources in the Earth-based network. Unlike other deci-Hz proposals (such as DECIGO), GLOC would continue to measure the \kj{heavy-stellar binaries} all the way up to merger and ringdown. This ensures that for stellar binaries, (i) the detection radius in in GLOC is comparable (or better) to the most futuristic concepts proposed for space-based detectors, (ii) the overall SNR in GLOC is about an order of magnitude higher than next-generation Earth-based detectors on Earth, and (ii) the effective baseline $L$ of GLOC would become comparable to the Moon's orbital diameter around the Earth. While this study sets the high-frequency sensitivity of GLOC $(\gtrsim 100$ Hz) to be comparable to CE, \kj{a future study will explore the advantage gained simply from the low-frequency improvement of GLOC where we find that most of the SNR is accumulated, as well as has the most impact on sky-localization.}

\begin{figure}
\centering
 \includegraphics[scale=0.42,trim = {95 20 30 40}]{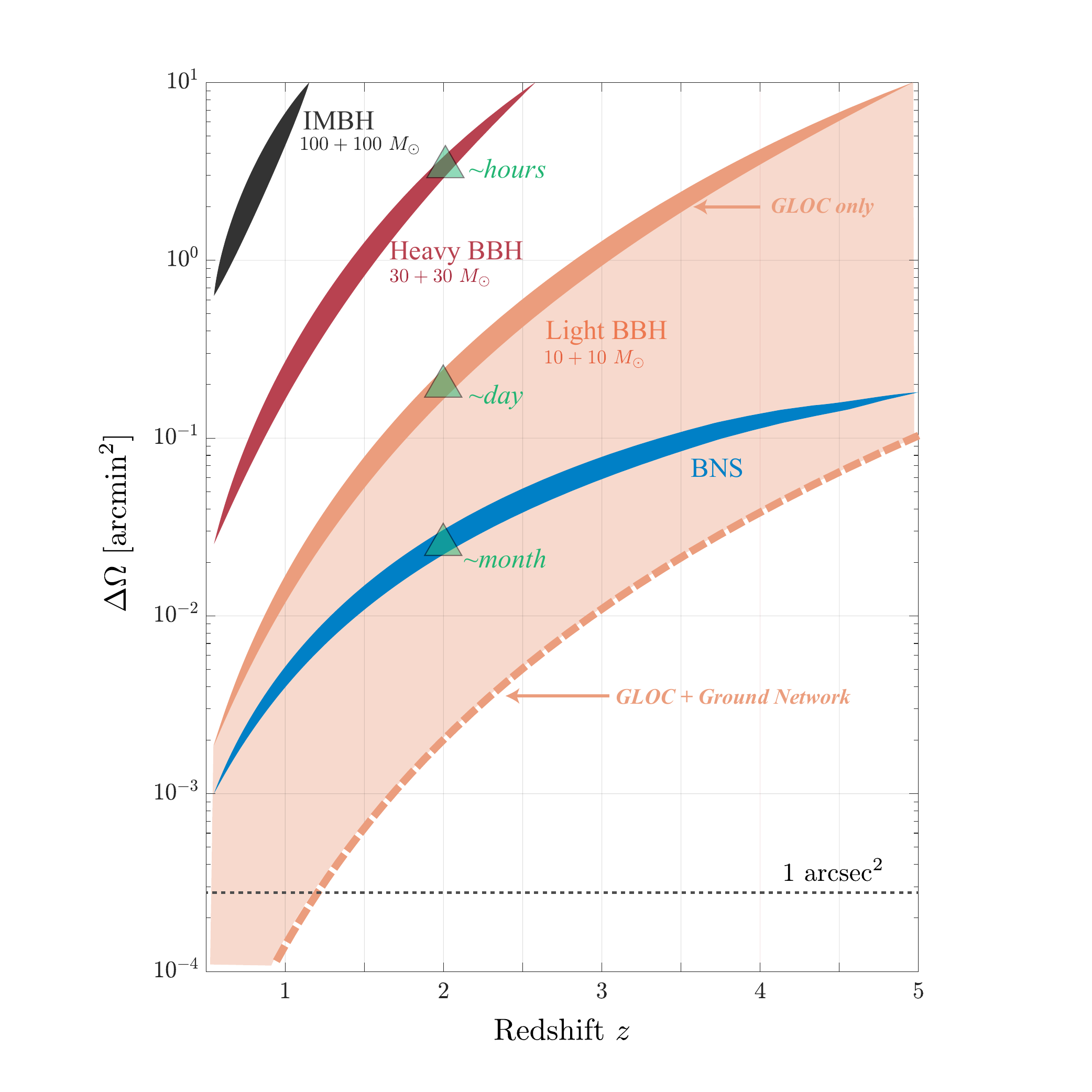}
\caption{{\bf Sky-localization with GLOC.} Each colored coded curve represents a face-on binary at different redshifts. The highlighted masses are in the source frame of the binary. The shaded region refers to the potential improvement for a $10+10~M_\odot$ binary with multiband network on GLOC (optimal) and next-generation Earth-based detectors.   }
\label{fig:sky} 
\end{figure}

The typical SNRs in GLOC for stellar binaries from high redshifts ($z\sim3-10)$ would be ${\sim} 100$, thus the timing accuracy, $\sigma_t$, could be as good as ${\sim} 0.1$ milli-seconds \cite{Fairhurst_2011}. This combined with the motion of the Moon at ${\sim} 1$~km/s will reduce the angular uncertainty to $\sigma_\theta \sim 10^{-3}~\mathrm{deg}$. As a result, the sky-localization of stellar binaries from GLOC alone will be $\Delta{\Omega}\sim 10^{-5}~\mathrm{deg}^2$. 
In Fig. \ref{fig:sky}, we show this approximate constraint on sky-localization for coalescing binaries at different redshifts.

A binary neutron star (BNS) at $z\sim2$ would be in the GLOC  band for an entire orbital period of the Moon, while a nearby BNS (like GW170817 \cite{GW170817bns}) would be in-band for almost three months. The sky-localization alert for BNSs can be sent days in advance, allowing readiness of high-latency electromagnetic followups with reach up to high redshifts.  {Even in the case of GLOC-conservative, the effective baseline for a BNS would be a quarter of the Moon's orbit, leading to tight constraints.}

For a relatively light binary black hole (BBH) like GW151226 \cite{GW151226bbh}, GLOC would start measuring its inspiral a day before the merger. These could constrain the sources at redshifts ${\sim}2$ to about $0.1~\mathrm{arcmin}^2$. Next generation Earth-based detectors CE and ET have their peak sensitivity for BBHs of these total masses \cite{2020NatAs...4..260J}. A combined network between these enhanced detectors on Earth and a geocentric detector like GLOC can further reduce the sky-location error by two orders of magnitude (see calculations in \cite{2020arXiv200401434G}). As shown in Fig. \ref{fig:sky} (dotted line), these combined network can constrain mergers of light BBHs to $1~\mathrm{arcsec}^2$, namely the angular scale of a single galaxy. These are the tightest constraints on the source location in \gw{} astronomy, allowing to identify the potential host galaxy without electromagnetic counterparts \cite{2016PhRvD..94b3516R, 2018JCAP...09..039S, 2020arXiv200708534S, 2020PhRvR...2b3314C}. The strongest science case of GLOC is opening such high redshift dark sirens to independently measure the evolution of the Hubble parameter as a function of redshift \cite{1986Natur.323..310S}.

For the emerging population of BBHs in the pair-instability supernovae (PISN) mass-gap \cite{Woosley:2002zz}, such as GW190521 \cite{Abbott:2020tfl, Abbott:2020mjq} and GW170502 \cite{Udall_2020}, GLOC would measure their inspiral a few hours before their mergers. These are the brightest sources in GLOC, registering SNR $\sim 100$ at $z\lesssim 20$. For comparison, a deci-Hz space-mission like DO would have similar reach only up to $z\lesssim 1$, while ET it can reach $z\lesssim 10$  (see fig. \ref{fig:hor}). A multiband observation between GLOC and ET would constrain the sky-location to ${\sim}1~\mathrm{arcmin}^2$, proving critical for associating flares in Active Galactic Nuclei (AGN) with such binaries \cite{PhysRevLett.124.251102}. Furthermore, the inclination angle and orbital eccentricity  will be tighter constrained by measuring the early inspiral in GLOC, thus reducing the error in redshift and the corresponding source-frame mass of black holes. The constraints on the secondary black hole will be particularly important in probing the lower-end of the PISN mass-gap. While a space-mission like LISA can measure the early-inspiral of these PISN BBHs years in advance \cite{2016PhRvL.116w1102S}, it can only do so for sources typically within a Gpc \cite{2020NatAs...4..260J} and retrospective after a detection from Earth \cite{Wong_2018}.

The enhanced low-frequency sensitivity permits GLOC to survey binaries with IMBHs of $10^{2-4}\ M_\odot$, practically across the entire universe. Such cosmological reach is crucial for connecting IMBHs with the Pop-III remnants \cite{Madau:2001sc} and the seeds of super-massive black holes \cite{Gair:2009gr, 2017A&G....58c3.22S, Pacucci_2020}. The theoretical estimates on their populations are fairly weak, but the upper-limits from LIGO and Virgo detectors \cite{O2IMBH} suggests that the mergers of lower-range IMBH binaries are much rarer ($\lesssim 1~\mathrm{Gpc}^{-3}~\mathrm{yr}^{-1}$) compared to stellar binaries. Unlike GLOC, space-missions have just a few years of lifetime ($4-10$ years for LISA), allowing the detection of only a handful of such rare sources.

In the case of potential IMBH detection, the advent of GLOC opens a new possibility of multiband observations across three frequency bands - from early inspiral at milli-Hz (space), to late-inspiral at deci-Hz (Moon) and mergers/ringdown at ${\sim}10$ Hz (Earth). Such joint measurements of a source across three bands of frequency spectrum would provide the strongest tests of the black hole area theorem, and no-hair theorem \cite{2016PhRvL.117e1102V, 2019BAAS...51c.109C, Sedda_2020}. The morphology of gravitational-waves in multiband of LISA and GLOC could provide a direct probe of the environment surrounding these black holes \cite{2020arXiv201006056T}. Furthermore, the intermediate-mass-ratio inspirals (IMRIs) \cite{Amaro_Seoane_2018}, which are relatively weak sources in both LISA and the next-generation Earth detectors, can be surveyed in GLOC to $z\sim10$. For IMRIs within redshift $\lesssim 1$, GLOC would measure them with SNR $\sim 100$. This would increase the detection confidence of these sources in LISA and ET, thus improving their overall parameter estimation. 

The long-term detection sensitivity at $\mathcal{O}(1~\mathrm{Hz})$ is unique to GLOC. This frequency regime is crucial for probing the explosion mechanism of Type Ia supernovae (SNe) \cite{Mandel_2018}. For the the single degenerate channel of Type Ia, we expect the \gw{} emission at around 1~Hz \cite{PhysRevLett.106.201103, 2015PhRvD..92l4013S}. If the progenitors of Type Ia SNe are mergers of two white dwarfs (double degenerate channel), we expect GLOC to detect them up to ${\sim}2~\mathrm{Gpc}$. These detections will also tightly bound the unknown masses of the white dwarfs, thus reducing the error budget in calibrating the standard candles. 

At the low end of source masses, GLOC can put the tightest bounds on a putative population of sub-solar dark matter objects ($0.1-1\, M_\odot)$ \cite{Shandera_2018}. There are no known astrophysical phenomena that can create detectable \gw{s} at such low-masses, however, primordial black holes or dark matter within neutron star cores offer possible scenarios (see \cite{Abbott_2019DM} and references within). The low-frequency sensitivity of GLOC allows us to measure the dark matter density of such exotic objects to 30\% of the entire observable volume of the universe ($z\sim10$).

\section{Conclusion and Future Work}
In this study, we demonstrate concept for Gravitational-wave Lunar Observatory for Cosmology (GLOC) -- a first of its kind proposal in the \kj{gravitational-wave detection era}. We find that GLOC offers the strongest science case towards fundamental cosmology, with an unprecedented access to over $70\%$ volume of the observable universe. The detection sensitivity of GLOC near $0.1-5$ Hz frequencies further opens a window to astrophysical sources and dark sirens that no other planned telescope would have for the next several decades. 

Grand challenges remain in terms of feasibility and cost of constructing a large-scale physics facility on the moon. Much of this will require technology demonstration on several fronts that currently possesses indefinite timeline. Nevertheless, exploration of a gravitational-wave setup can be initiated through the first set of astronaut that will return to the moon this decade. In particular, monitoring of seismic activity near the deci-Hertz regime would prove crucial in getting a noise budget for GLOC. A prototype GLOC-pathfinder can be envisioned in the time scale of the several next-generation gravitational-wave experiments. In collaboration with other lunar-based concepts LGWA \cite{Lunar_Harms} and LSGA \cite{LSGA}, studies are currently being planned in further exploring the science case of multi-messenger astronomy from the moon and engaging with the wider community.

\section{Acknowledgements}
 {We are grateful to Rainer Weiss, Kelly Holley-Bockelmann, Stavros Katsanevas, Scott Hughes, Mathew Evans, Bob Eisenstein, Kevin Kuns, Brian O'Reilly, Stefan Hild, Jan Harms, Szabolcs Marka, Philippe Lognonné, Robin Stebbins, Peter Bender, Robert Fisher and the referees for insightful comments}. K.J’s research was supported by the GRAVITY program at Vanderbilt University. A. L.'s work was supported in part by the Black Hole Initiative at Harvard University, which is funded by grants from the John Templeton Foundation and the Gordon and Betty Moore Foundation.

%\bibliography{references.bib}
\bibliographystyle{iopart-num}
\bibliography{references}

\providecommand{\newblock}{}
\begin{thebibliography}{10}
\expandafter\ifx\csname url\endcsname\relax
  \def\url#1{{\tt #1}}\fi
\expandafter\ifx\csname urlprefix\endcsname\relax\def\urlprefix{URL }\fi
\providecommand{\eprint}[2][]{\url{#2}}

\bibitem{Abbott_2019}
Abbott B, Abbott R, Abbott T, Abraham S, Acernese F, Ackley K, Adams C,
  Adhikari R, Adya V, Affeldt C and et~al 2019 {\em Physical Review X\/} {\bf
  9} ISSN 2160-3308 \urlprefix\url{http://dx.doi.org/10.1103/PhysRevX.9.031040}

\bibitem{Punturo:2010zz}
Punturo M {\em et~al.\/} 2010 {\em Class. Quant. Grav.\/} {\bf 27} 194002

\bibitem{CosmicExplorer}
Reitze D {\em et~al.\/} 2019 {\em arXiv e-prints\/} arXiv:1907.04833
  (\textit{Preprint} \eprint{1907.04833})

\bibitem{LISAmain}
Amaro-Seoane P {\em et~al.\/} 2017 {\em arXiv e-prints\/} arXiv:1702.00786
  (\textit{Preprint} \eprint{1702.00786})

\bibitem{IPTA):2013lea}
Manchester R~N 2013 {\em Class. Quant. Grav.\/} {\bf 30} 224010
  (\textit{Preprint} \eprint{1309.7392})

\bibitem{Harry_2006}
Harry G~M, Fritschel P, Shaddock D~A, Folkner W and Phinney E~S 2006 {\em
  Classical and Quantum Gravity\/} {\bf 23} 4887--4894
  \urlprefix\url{https://doi.org/10.1088%2F0264-9381%2F23%2F15%2F008}

\bibitem{muARES}
Sesana A, Korsakova N, Sedda M~A {\em et~al.\/} 2019 Unveiling the
  gravitational universe at $\mu$-hz frequencies (\textit{Preprint}
  \eprint{1908.11391})

\bibitem{AMIGO}
Baibhav V, Barack L, Berti E {\em et~al.\/} 2019 Probing the nature of black
  holes: Deep in the mhz gravitational-wave sky (\textit{Preprint}
  \eprint{1908.11390})

\bibitem{Mandel_2018}
Mandel I, Sesana A and Vecchio A 2018 {\em Classical and Quantum Gravity\/}
  {\bf 35} 054004 ISSN 1361-6382
  \urlprefix\url{http://dx.doi.org/10.1088/1361-6382/aaa7e0}

\bibitem{Sedda_2020}
Sedda M~A, Berry C~P~L, Jani K {\em et~al.\/} 2020 {\em Classical and Quantum
  Gravity\/} {\bf 37} 215011
  \urlprefix\url{https://doi.org/10.1088%2F1361-6382%2Fabb5c1}

\bibitem{Lacour_2019}
Lacour S, Vincent F~H, Nowak M, Le~Tiec A, Lapeyrere V, David L, Bourget P,
  Kellerer A, Jani K, Martino J and et~al 2019 {\em Classical and Quantum
  Gravity\/} {\bf 36} 195005 ISSN 1361-6382
  \urlprefix\url{http://dx.doi.org/10.1088/1361-6382/ab3583}

\bibitem{2019IJMPD..2845001K}
{Kawamura} S {\em et~al.\/} 2019 {\em International Journal of Modern Physics
  D\/} {\bf 28} 1845001

\bibitem{Crowder_2005}
Crowder J and Cornish N~J 2005 {\em Physical Review D\/} {\bf 72} ISSN
  1550-2368 \urlprefix\url{http://dx.doi.org/10.1103/PhysRevD.72.083005}

\bibitem{alma991023959399703276}
Thorne K~S {\em Black holes and time warps : Einstein's outrageous legacy\/}
  (W.W. Norton) ISBN 0393035050

\bibitem{1993LPI....24..841L}
{Lafave} N and {Wilson} T~L 1993 {Lunar LIGO: A New Concept in Gravitational
  Wave Astronomy} {\em Lunar and Planetary Science Conference\/} Lunar and
  Planetary Science Conference p 841

\bibitem{potter_physics_1990}
Potter A~E and Wilson T~L (eds) 1990 {\em Physics and astrophysics from a lunar
  base: first {NASA} workshop, {Stanford}, {CA}, 1989\/} ({\em {AIP} conference
  proceedings\/} no 202) (New York: American Institute of Physics) ISBN
  9780883186466

\bibitem{AdvLIGORef}
{{LIGO Scientific Collaboration}} 2015 {\em Classical and Quantum Gravity\/}
  {\bf 32} 074001 (\textit{Preprint} \eprint{1411.4547})

\bibitem{Johnson1972}
Johnson F~S, Carroll J~M and Evans D~E 1972 {\em Journal of Vacuum Science and
  Technology\/} {\bf 9} 450--456
  \urlprefix\url{https://doi.org/10.1116/1.1316652}

\bibitem{Larose2005}
Larose E 2005 {\em Geophysical Research Letters\/} {\bf 32}
  \urlprefix\url{https://doi.org/10.1029/2005gl023518}

\bibitem{Philippe_moon_seis}
Lognonné P, Le~Feuvre M, Johnson C~L and Weber R~C 2009 {\em Journal of
  Geophysical Research: Planets\/} {\bf 114}
  \urlprefix\url{https://agupubs.onlinelibrary.wiley.com/doi/abs/10.1029/2008JE003294}

\bibitem{Hanada2005}
Hanada H, Heki K, Araki H, Matsumoto K, Noda H, Kawano N, Tsubokawa T, Tsuruta
  S, Tazawa S, Asari K, Kono Y, Yano T, Gouda N, Iwata T, Yokoyama T, Kanamori
  H, Funazaki K and Miyazaki T 2005 Application of a {PZT} telescope to in situ
  lunar orientation measurement ({ILOM}) {\em International Association of
  Geodesy Symposia\/} (Springer Berlin Heidelberg) pp 163--168
  \urlprefix\url{https://doi.org/10.1007/3-540-27432-4_29}

\bibitem{Evans:2016mbw}
Abbott B~P {\em et~al.\/} (LIGO Scientific) 2017 {\em Class. Quant. Grav.\/}
  {\bf 34} 044001 (\textit{Preprint} \eprint{1607.08697})

\bibitem{Goda2008}
Goda K, Miyakawa O, Mikhailov E~E, Saraf S, Adhikari R, McKenzie K, Ward R,
  Vass S, Weinstein A~J and Mavalvala N 2008 {\em Nature Physics\/} {\bf 4}
  472--476 \urlprefix\url{https://doi.org/10.1038/nphys920}

\bibitem{Hild_2009}
Hild S, Chelkowski S, Freise A, Franc J, Morgado N, Flaminio R and DeSalvo R
  2009 {\em Classical and Quantum Gravity\/} {\bf 27} 015003 ISSN 1361-6382
  \urlprefix\url{http://dx.doi.org/10.1088/0264-9381/27/1/015003}

\bibitem{Hild_2011}
Hild S, Abernathy M, Acernese F, Amaro-Seoane P, Andersson N, Arun K, Barone F,
  Barr B, Barsuglia M, Beker M and et~al 2011 {\em Classical and Quantum
  Gravity\/} {\bf 28} 094013 ISSN 1361-6382
  \urlprefix\url{http://dx.doi.org/10.1088/0264-9381/28/9/094013}

\bibitem{Williams2017}
Williams J~P, Paige D, Greenhagen B and Sefton-Nash E 2017 {\em Icarus\/} {\bf
  283} 300--325 \urlprefix\url{https://doi.org/10.1016/j.icarus.2016.08.012}

\bibitem{Chappaz2017}
Chappaz L, Sood R, Melosh H~J, Howell K~C, Blair D~M, Milbury C and Zuber M~T
  2017 {\em Geophysical Research Letters\/} {\bf 44} 105--112
  \urlprefix\url{https://doi.org/10.1002/2016gl071588}

\bibitem{Robson_2019}
Robson T, Cornish N~J and Liu C 2019 {\em Classical and Quantum Gravity\/} {\bf
  36} 105011 ISSN 1361-6382
  \urlprefix\url{http://dx.doi.org/10.1088/1361-6382/ab1101}

\bibitem{PhysRevLett.122.231102}
Craig K, Steinlechner J, Murray P~G, Bell A~S, Birney R, Haughian K, Hough J,
  MacLaren I, Penn S, Reid S, Robie R, Rowan S and Martin I~W 2019 {\em Phys.
  Rev. Lett.\/} {\bf 122}(23) 231102
  \urlprefix\url{https://link.aps.org/doi/10.1103/PhysRevLett.122.231102}

\bibitem{2014CQGra..31x5006F}
{Friedrich} D, {Nakano} M, {Kawamura} H, {Yamanaka} Y, {Hirobayashi} S and
  {Kawamura} S 2014 {\em Classical and Quantum Gravity\/} {\bf 31} 245006

\bibitem{pygwinc}
GWINC 2020 pygwinc \url{https://git.ligo.org/gwinc/pygwinc}

\bibitem{2020NatAs...4..260J}
{Jani} K, {Shoemaker} D and {Cutler} C 2019 {\em Nature Astronomy\/} {\bf 4}
  260--265 (\textit{Preprint} \eprint{1908.04985})

\bibitem{London_2018}
London L, Khan S, Fauchon-Jones E, García C, Hannam M, Husa S,
  Jiménez-Forteza X, Kalaghatgi C, Ohme F and Pannarale F 2018 {\em Physical
  Review Letters\/} {\bf 120} ISSN 1079-7114
  \urlprefix\url{http://dx.doi.org/10.1103/PhysRevLett.120.161102}

\bibitem{PhysRevD.97.024016}
Calder\'on~Bustillo J, Salemi F, Dal~Canton T and Jani K~P 2018 {\em Phys. Rev.
  D\/} {\bf 97}(2) 024016
  \urlprefix\url{https://link.aps.org/doi/10.1103/PhysRevD.97.024016}

\bibitem{2018arXiv180706209P}
{Planck Collaboration} 2018 {\em arXiv e-prints\/} arXiv:1807.06209
  (\textit{Preprint} \eprint{1807.06209})

\bibitem{Fairhurst_2011}
Fairhurst S 2011 {\em Classical and Quantum Gravity\/} {\bf 28} 105021 ISSN
  1361-6382 \urlprefix\url{http://dx.doi.org/10.1088/0264-9381/28/10/105021}

\bibitem{Lau_2020}
Lau M~Y~M, Mandel I, Vigna-Gómez A, Neijssel C~J, Stevenson S and Sesana A
  2020 {\em Monthly Notices of the Royal Astronomical Society\/} {\bf 492}
  3061–3072 ISSN 1365-2966
  \urlprefix\url{http://dx.doi.org/10.1093/mnras/staa002}

\bibitem{ObservingScenarios}
{The LIGO Scientific Collaboration} and {the Virgo Collaboration} 2013 {\em
  arXiv:1304.0670\/} (\textit{Preprint} \eprint{1304.0670})

\bibitem{Shandera_2018}
Shandera S, Jeong D and Gebhardt H~S~G 2018 {\em Physical Review Letters\/}
  {\bf 120} ISSN 1079-7114
  \urlprefix\url{http://dx.doi.org/10.1103/PhysRevLett.120.241102}

\bibitem{2006ApJ...653L..53A}
{Amaro-Seoane} P and {Freitag} M 2006 {\em Astrophysical Journal Letters\/}
  {\bf 653} L53--L56

\bibitem{burns2012astrophysics}
Burns J~O, Lazio T~J~W and Bottke W 2012 Astrophysics conducted by the lunar
  university network for astrophysics research (lunar) and the center for lunar
  origins (cloe) (\textit{Preprint} \eprint{1209.2233})

\bibitem{Avibook}
Loeb A and Furlanetto S~R 2013 {\em The First Galaxies in the Universe\/} stu -
  student edition ed (Princeton University Press) ISBN 9780691144917
  \urlprefix\url{http://www.jstor.org/stable/j.ctt24hrpv}

\bibitem{Koushiappas_2017}
Koushiappas S~M and Loeb A 2017 {\em Physical Review Letters\/} {\bf 119} ISSN
  1079-7114 \urlprefix\url{http://dx.doi.org/10.1103/PhysRevLett.119.221104}

\bibitem{Jani_2020}
Jani K and Loeb A 2020 {\em The Astrophysical Journal\/} {\bf 889} L35
  \urlprefix\url{https://doi.org/10.3847\%2F2041-8213\%2Fab6854}

\bibitem{Peters1964}
{Peters} P~C 1964 {\em Physical Review\/} {\bf 136} 1224--1232

\bibitem{GW170817bns}
Collaboration L~S and Collaboration V 2017 {\em Phys. Rev. Lett.\/} {\bf
  119}(16) 161101
  \urlprefix\url{https://link.aps.org/doi/10.1103/PhysRevLett.119.161101}

\bibitem{GW151226bbh}
Collaboration L~S and Collaboration V 2016 {\em Physical Review Letters\/} {\bf
  116} 241103 (\textit{Preprint} \eprint{1606.04855})

\bibitem{2020arXiv200401434G}
{Grimm} S and {Harms} J 2020 {\em arXiv e-prints\/} arXiv:2004.01434
  (\textit{Preprint} \eprint{2004.01434})

\bibitem{2016PhRvD..94b3516R}
{Raccanelli} A, {Kovetz} E~D, {Bird} S, {Cholis} I and {Mu{\~n}oz} J~B 2016
  {\em Phys. Rev. D\/} {\bf 94} 023516 (\textit{Preprint} \eprint{1605.01405})

\bibitem{2018JCAP...09..039S}
{Scelfo} G, {Bellomo} N, {Raccanelli} A, {Matarrese} S and {Verde} L 2018 {\em
  J. Cosmol. Astropart. Phys.\/} {\bf 2018} 039 (\textit{Preprint}
  \eprint{1809.03528})

\bibitem{2020arXiv200708534S}
{Scelfo} G, {Boco} L, {Lapi} A and {Viel} M 2020 {\em arXiv e-prints\/}
  arXiv:2007.08534 (\textit{Preprint} \eprint{2007.08534})

\bibitem{2020PhRvR...2b3314C}
{Calore} F, {Cuoco} A, {Regimbau} T, {Sachdev} S and {Serpico} P~D 2020 {\em
  Physical Review Research\/} {\bf 2} 023314 (\textit{Preprint}
  \eprint{2002.02466})

\bibitem{1986Natur.323..310S}
{Schutz} B~F 1986 {\em Nature\/} {\bf 323} 310--311

\bibitem{Woosley:2002zz}
Woosley S~E, Heger A and Weaver T~A 2002 {\em Rev. Mod. Phys.\/} {\bf 74}
  1015--1071

\bibitem{Abbott:2020tfl}
Abbott R {\em et~al.\/} (LIGO Scientific, Virgo) 2020 {\em Phys. Rev. Lett.\/}
  {\bf 125} 101102 (\textit{Preprint} \eprint{2009.01075})

\bibitem{Abbott:2020mjq}
Abbott R {\em et~al.\/} (LIGO Scientific, Virgo) 2020 {\em Astrophys. J.
  Lett.\/} {\bf 900} L13 (\textit{Preprint} \eprint{2009.01190})

\bibitem{Udall_2020}
Udall R {\em et~al.\/} 2020 {\em The Astrophysical Journal\/} {\bf 900} 80
  \urlprefix\url{https://doi.org/10.3847%2F1538-4357%2Fabab9d}

\bibitem{PhysRevLett.124.251102}
Graham M~J {\em et~al.\/} 2020 {\em Phys. Rev. Lett.\/} {\bf 124}(25) 251102
  \urlprefix\url{https://link.aps.org/doi/10.1103/PhysRevLett.124.251102}

\bibitem{2016PhRvL.116w1102S}
{Sesana} A 2016 {\em Physical Review Letters\/} {\bf 116} 231102

\bibitem{Wong_2018}
Wong K~W, Kovetz E~D, Cutler C and Berti E 2018 {\em Physical Review Letters\/}
  {\bf 121} ISSN 1079-7114
  \urlprefix\url{http://dx.doi.org/10.1103/PhysRevLett.121.251102}

\bibitem{Madau:2001sc}
Madau P and Rees M~J 2001 {\em Astrophys. J.\/} {\bf 551} L27--L30
  (\textit{Preprint} \eprint{astro-ph/0101223})

\bibitem{Gair:2009gr}
Gair J~R, Mandel I, Sesana A and Vecchio A 2009 {\em Class. Quant. Grav.\/}
  {\bf 26} 204009 (\textit{Preprint} \eprint{0907.3292})

\bibitem{2017A&G....58c3.22S}
{Smith} A, {Bromm} V and {Loeb} A 2017 {\em Astronomy and Geophysics\/} {\bf
  58} 3.22--3.26 (\textit{Preprint} \eprint{1703.03083})

\bibitem{Pacucci_2020}
Pacucci F and Loeb A 2020 {\em The Astrophysical Journal\/} {\bf 895} 95 ISSN
  1538-4357 \urlprefix\url{http://dx.doi.org/10.3847/1538-4357/ab886e}

\bibitem{O2IMBH}
{The LIGO Scientific Collaboration} and {the Virgo Collaboration} 2019 {\em
  Physical Review D\/} {\bf 100} ISSN 2470-0029
  \urlprefix\url{http://dx.doi.org/10.1103/PhysRevD.100.064064}

\bibitem{2016PhRvL.117e1102V}
{Vitale} S 2016 {\em Physical Review Letters\/} {\bf 117} 051102

\bibitem{2019BAAS...51c.109C}
{Cutler} C, {Berti} E, {Holley-Bockelmann} K, {Jani} K, {Kovetz} E~D, {Larson}
  S~L, {Littenberg} T, {McWilliams} S~T, {Mueller} G, {Randall} L, {Schnittman}
  J~D, {Shoemaker} D~H, {Vallisneri} M, {Vitale} S and {Wong} K~W~K 2019 {\em
  Bull. Am. Astron. Soc.\/} {\bf 51} 109 (\textit{Preprint}
  \eprint{1903.04069})

\bibitem{2020arXiv201006056T}
{Toubiana} A {\em et~al.\/} 2020 {\em arXiv e-prints\/} arXiv:2010.06056
  (\textit{Preprint} \eprint{2010.06056})

\bibitem{Amaro_Seoane_2018}
Amaro-Seoane P 2018 {\em Physical Review D\/} {\bf 98} ISSN 2470-0029
  \urlprefix\url{http://dx.doi.org/10.1103/PhysRevD.98.063018}

\bibitem{PhysRevLett.106.201103}
Falta D, Fisher R and Khanna G 2011 {\em Phys. Rev. Lett.\/} {\bf 106}(20)
  201103
  \urlprefix\url{https://link.aps.org/doi/10.1103/PhysRevLett.106.201103}

\bibitem{2015PhRvD..92l4013S}
{Seitenzahl} I~R, {Herzog} M, {Ruiter} A~J, {Marquardt} K, {Ohlmann} S~T and
  {R{\"o}pke} F~K 2015 {\em Phys. Rev. D\/} {\bf 92} 124013 (\textit{Preprint}
  \eprint{1511.02542})

\bibitem{Abbott_2019DM}
Abbott B, Abbott R, Abbott T, Abraham S, Acernese F, Ackley K, Adams C,
  Adhikari R, Adya V, Affeldt C and et~al 2019 {\em Physical Review Letters\/}
  {\bf 123} ISSN 1079-7114
  \urlprefix\url{http://dx.doi.org/10.1103/PhysRevLett.123.161102}

\bibitem{Lunar_Harms}
{Harms} J {\em et~al.\/} 2020 {\em arXiv e-prints\/} arXiv:2010.13726
  (\textit{Preprint} \eprint{2010.13726})

\bibitem{LSGA}
Katsanevas S {\em et~al.\/} 2020 {\em Ideas for exploring the Moon with a large
  European lander (ESA)\/}
  \urlprefix\url{https://ideas.esa.int/servlet/hype/IMT?documentTableId=45087607031744010&userAction=Browse&templateName=&documentId=a315450fae481074411ef65e4c5b7746}

\end{thebibliography}

%\end{thebibliography}
\end{document}